\documentclass[pra,a4paper,aps,twocolumn,superscriptaddress,longbibliography,10pt]{revtex4-2}

\usepackage{amsmath, empheq,braket,amsthm,amssymb,graphics,amsfonts,array,color,subfigure, mathrsfs}
\usepackage{latexsym}
\usepackage{tensor}
\usepackage{empheq}

\usepackage[unicode=true,pdfusetitle, bookmarks=true,bookmarksnumbered=false,bookmarksopen=false,breaklinks=false,pdfborder={0 0 0},backref=false,colorlinks=true,linkcolor=blue,citecolor=blue,urlcolor=blue]{hyperref}

\newcommand{\proj}[1]{\ket{#1}\bra{#1}}

\newcommand{\ad}{\hat a^{\dagger}}
\newcommand{\bd}{\hat b^{\dagger}}

\newcommand{\bs}{U^{\mathrm{BS}}_{\eta}}
\newcommand{\bsd}{U^{\mathrm{BS} \dagger}_{\eta}}
\newcommand{\tms}{U^{\mathrm{TMS}}_{\lambda}}
\newcommand{\tmsd}{U^{\mathrm{TMS} \dagger}_{\lambda}}

\newtheorem{theo}{Theorem}

\newcommand{\zt}[3]{\mathcal{T}_{#1}\left[#2\right]\left(#3\right)}

\renewcommand{\a}{\hat{a}}
\renewcommand{\b}{\hat{b}}

\newcommand{\eqal}[1]{\begin{equation}
\begin{aligned}
#1
\end{aligned}
\end{equation}}

\newcommand{\eqaln}[1]{\begin{equation} \nonumber
\begin{aligned}
#1
\end{aligned}
\end{equation}}

\begin{document}


\title{Multiparticle quantum interference in Bogoliubov bosonic transformations}

\author{Michael G. Jabbour}
\email{mgija@dtu.dk}
\affiliation{Quantum Information and Communication, \'Ecole polytechnique de Bruxelles, CP 165/59, Universit\'e libre de Bruxelles, 1050 Brussels, Belgium}
\affiliation{Department of Applied Mathematics and Theoretical Physics, University of Cambridge, Cambridge CB3 0WA, United Kingdom}
\affiliation{Department of Physics, Technical University of Denmark, 2800 Kongens Lyngby, Denmark}
\author{Nicolas J. Cerf}
\email{ncerf@ulb.ac.be}
\affiliation{Quantum Information and Communication, \'Ecole polytechnique de Bruxelles, CP 165/59, Universit\'e libre de Bruxelles, 1050 Brussels, Belgium}


\begin{abstract}
We explore the multiparticle transition probabilities in Gaussian unitaries effected by a two-mode Bogoliubov bosonic transformation on the mode annihilation and creation operators. We show that the transition probabilities can be characterized by remarkably simple, yet unsuspected recurrence equations involving a linear combination of probabilities. The recurrence exhibits an interferometric suppression term -- a \textit{negative} probability -- which generalizes the seminal Hong-Ou-Mandel effect to more than two indistinguishable photons impinging on a beam splitter of rational transmittance. Unexpectedly, interferences thus originate in this description from the cancellation of probabilities instead of amplitudes. Our framework, which builds on the generating function of the non-Gaussian matrix elements of Gaussian unitaries in Fock basis, is illustrated here for the most common passive and active linear coupling between two optical modes driven by a beam splitter or a parametric amplifier. Hence, it also allows us to predict unsuspected multiphoton interference effects in an optical amplifier of rational gain. In particular, we confirm the newly found two-photon interferometric suppression effect in an amplifier of gain 2 originating from timelike indistinguishability \href{https://doi.org/10.1073/pnas.2010827117}{[Proc. Natl. Acad. Sci. \textbf{117}, 33107 (2020)]}. Overall, going beyond standard two-mode optical components, we expect our method will prove valuable for describing general quantum circuits involving Bogoliubov bosonic transformations.
\end{abstract}

\maketitle

\section{Introduction}
Quantum interference is a cornerstone of quantum physics. While it challenges our understanding of the universe as for instance witnessed in Young's celebrated double slit-experiment, it has various applications such as quantum computing \cite{Ladd2010}, quantum cryptography \cite{Gisin2002}, or superconducting quantum interference devices \cite{Vasyukov2013}. Quantum interference is notably a key to implementing future quantum technologies with photonic integrated devices, which has resulted in a vigorous research effort on harnessing multimode multiphoton interferences over the last decade, see e.g.~\cite{Peruzzo2011,Walmsley2019}. This is also significant in connection with the boson sampling paradigm~\cite{bosonsampling}, which builds on the computational hardness of simulating the coherent propagation of many identical bosons through a multimode linear interferometer and holds the promise of substantiating the advantage of quantum computers~\cite{Tillmann2013,Crespi2013,Broome2013,Carolan2014}. More generally, this has led to a revived interest for quantum interferometry going beyond the celebrated Hong-Ou-Mandel (HOM) effect~\cite{HOM}, e.g., the generalized bunching effect in linear networks~\cite{Shchesnovich2016}, the signatures of nonclassicality in a multimode interferometer~\cite{Rigovacca2016}, the observation of intrinsically 3-photon interference~\cite{Agne2017,Menssen2017}, or the suppression laws in a 8-mode optical Fourier interferometer~\cite{Crespi2016}.
\par

Formally, quantum interferences originate from adding up the amplitudes of (often a large number of) possible paths.  Since amplitudes are complex,  taking the square modulus of the resulting sum typically gives rise to constructive or destructive interferences. The HOM effect is a paradigmatic example of two-photon quantum interference: the probability of detecting two photons in coincidence at the output of a 50:50 beam splitter (one in each mode) vanishes when one photon impinges on each of the two input modes. The sum of the amplitudes of the two possible paths (both photons being either reflected or transmitted) vanishes, giving rise to destructive interference. 
In a nutshell, when only two paths of amplitudes $\alpha_1$ and $\alpha_2$ interfere, the resulting probability is $p=|\alpha_1+\alpha_2|^2=p_1+p_2+2  \sqrt{p_1 p_2} \cos\theta$, where  $p_1=|\alpha_1|^2$, $p_2=|\alpha_2|^2$, and $\theta$ is the relative phase.

In this paper, we explore multiparticle quantum interferences that emerge in Bogoliubov bosonic transformations. Bogoliubov transformations are ubiquitous in physics, appearing in various fields such as superconductivity, superfluidity, nuclear physics and quantum field theory. They are also essential in understanding phenomena such as Hawking radiation \cite{Hawking1975,Traschen1999_3} and the Unruh effect \cite{birrell_davies_1982,Crispino2008}. While our methods and results could be applied in various situations involving bosonic systems, we choose to illustrate them here by focusing on the quantum optics framework. Specifically, we investigate the optical Gaussian unitaries effected by Bogoliubov transformations in phase space, which closely model a great amount of operations performed in quantum optics experiments \cite{Weedbrook2011}. We start by examining the generic case of $i$ and $k$ photons impinging on the two input modes of a beam splitter, one of the simplest yet most essential operations in any optical setting. The probability $B_{n}^{(i,k)}$ of any output pattern is known to be expressible as  a multiple summation involving four binomial coefficients [see Eq.~\eqref{eq:complicated}]. This complicated expression, owing to the many interfering paths, can of course be evaluated but cannot easily be exploited analytically. Here, we derive an unexpectedly simple formula (Theorem~\ref{theo:mainBS}) which governs these probabilities. Counterintuitively, it involves a simple linear combination of probabilities (with no usual $\sqrt{p_1 p_2} $ terms) and the discrepancy with respect to the corresponding classical formula for distinguishable photons appears as a \textit{negative} probability \cite{Feynman}. 

Our technique relies on calculating the generating function of the matrix elements of Gaussian unitaries in Fock basis, which can be expressed in a simple closed form with the Gaussian toolbox although it encapsulates complex non-Gaussian features such as the multiphoton transition probabilities $B_{n}^{(i,k)}\!$. It allows us to extend the HOM effect to many photons by predicting a simple negative contribution to the transition probability. More generally, our framework is suited to Gaussian unitaries describing the passive but also the active linear coupling between two bosonic modes. Hence, we predict a similar interference suppression term -- a negative probability -- in an optical amplifier, such as a nonlinear crystal pumped in the nondegenerate parametric amplification regime or a four-wave mixer (see Theorem~\ref{theo:mainTMS}). This corroborates and extends the recent finding of a two-boson interference effect in a gain-2 amplifier originating from timelike indistinguishability (bosons from the past and future cannot be distinguished) \cite{HOM_PTR_2}. Active optical components are essential in continuous-variable quantum information processing \cite{RevModPhys.77.513,CLP2007} as they give access to invaluable resources and protocols, such as universal computing with Gaussian cluster states \cite{PhysRevA.76.032321,PhysRevA.79.062318,Larsen369}, optical multimode entanglement \cite{PhysRevLett.114.050501}, Gaussian quantum steering \cite{PhysRevLett.114.060403}, or Gaussian quantum cloning~\cite{cloning}.

As a last result, we provide a further generalization of the HOM effect and two-boson active interference effect~\cite{HOM_PTR_2} by predicting a full interferometric suppression for any rational value of the transmittance (or gain) of a passive (or active) transformation provided specific photon numbers are chosen. Furthermore, we also briefly show that the asymptotic behavior of interferences with large photon numbers can easily be accessed based on generating functions. This illustrates the potential of our framework for describing multiparticle interferences in quantum circuits involving bosonic Bogoliubov transformations in phase space.


\section {Model and derivations}
\subsection{Bosonic Gaussian unitaries}
Bosonic modes are common carriers of continuous-variable quantum information~\cite{RevModPhys.77.513,CLP2007}. A bosonic mode (e.g., a quantized mode of the electromagnetic field) is modelled by a quantum harmonic oscillator in an infinite-dimensional Fock space. It is associated with the usual pair of bosonic mode operators $\hat a$ and $\ad$, which must satisfy the commutation relation $[\hat a,\ad]=\mathbb{I}$. In this context, Bogoliubov transformations~\cite{Bogoliubov} (i.e., linear canonical transformations in $\hat a$ and $\ad$) are of particular interest as they correspond to Gaussian unitaries (i.e., quadratic Hamiltonians in $\hat a$ and $\ad$). They are especially valuable in the framework of quantum optics, where they conserve Gaussian-shaped Wigner functions in phase space and, most importantly, model ubiquitous transformations in experimental conditions and form the core of Gaussian quantum information~\cite{Weedbrook2011}. They can be divided into passive and active transformations as effected by linear optical interferometry or parametric amplification, respectively.  In this work, we illustrate our method for the most fundamental passive and active two-mode Gaussian unitaries, namely the beam splitter (BS) and two-mode squeezer (TMS). The BS unitary $\bs$ effects an energy-conserving linear coupling between two modes and acts in the Heisenberg picture as
\begin{equation} \label{eq:BogoliubovBS}
\begin{aligned}
\bsd \, \hat a \, \bs & = \sqrt{\eta} \, \hat a + \sqrt{1-\eta} \, \hat b, \\
\bsd \, \hat b \, \bs & = - \sqrt{1-\eta} \, \hat a + \sqrt{\eta} \, \hat b,
\end{aligned}
\end{equation}
where $\hat a$ and $\hat b$ are the mode operators, while $\eta$ is the transmittance. Similarly, the TMS unitary $ \tms$ models the generation of pairs of entangled photons by parametric amplification due to the pumping of a nonlinear crystal, and acts on mode operators as
\begin{equation} \label{eq:BogoliubovTMS}
\begin{aligned}
\tmsd \, \hat a \, \tms & = \cosh(r) \, \hat a + \sinh(r) \, \bd, \\
\tmsd \, \hat b \, \tms & = \sinh(r) \, \ad + \cosh(r) \, \hat b.
\end{aligned}
\end{equation}
with $\lambda \coloneqq \tanh^2 (r)$ for a parametric gain $g  \coloneqq \cosh^2(r)$. The transformations characterized by Eqs. \eqref{eq:BogoliubovBS} and \eqref{eq:BogoliubovTMS} happen to be useful in various contexts involving the evolution of bosonic systems. For instance, they can be exploited in black hole theory, where they describe the interaction of a Gaussian bosonic state with an already formed Schwarzchild black hole \cite{AdamiBlackHoles}.
\par

\subsection{Generating functions}

\begin{figure}[t]
	\includegraphics[trim = 0cm 0cm 0cm 0cm, clip, width=0.6\columnwidth]{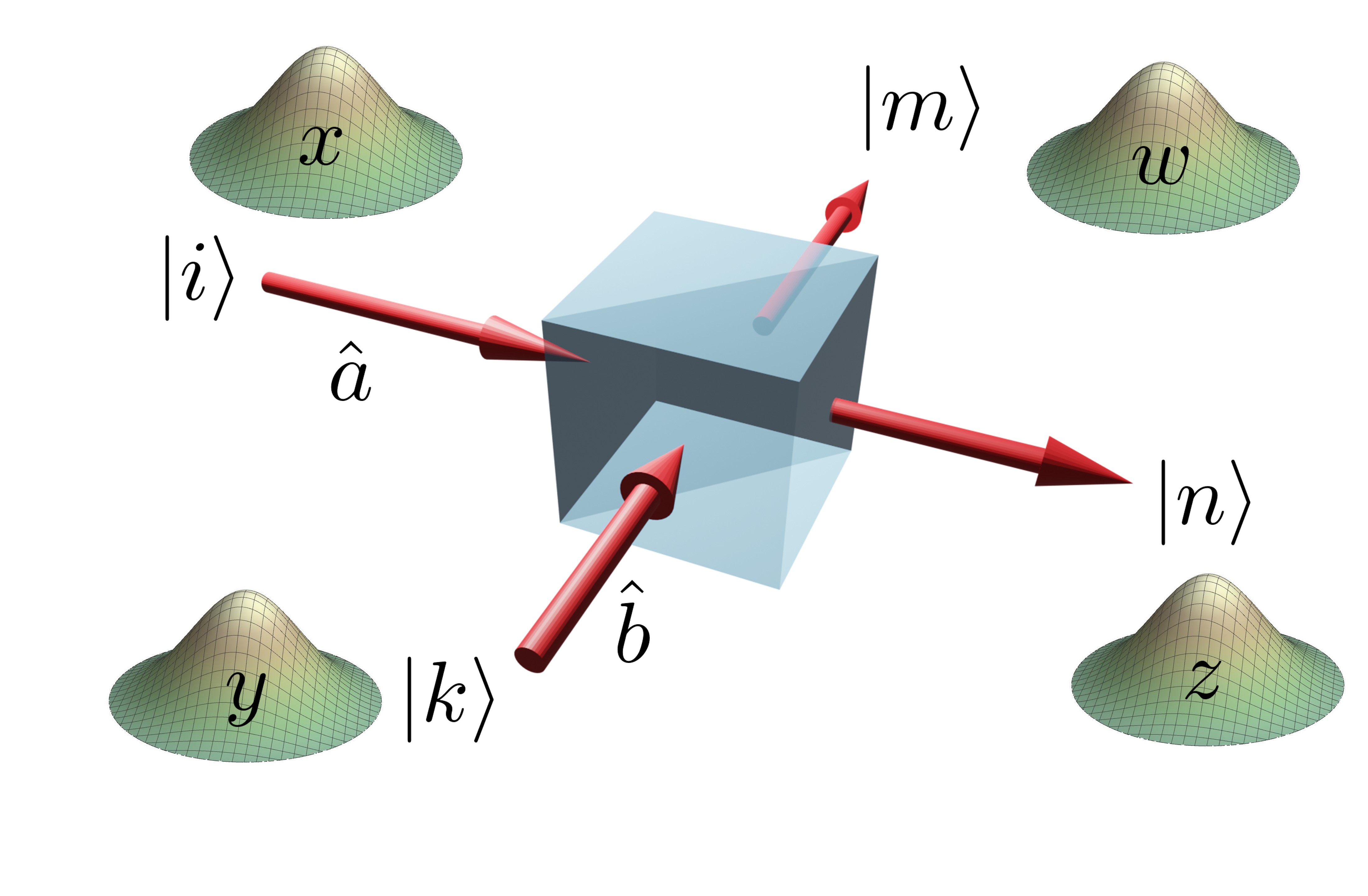}
	\caption{\label{conventions} Conventions in the definition of $f(x,y,z,w)$, which is the generating function of the transition probability for sending the Fock states $\ket{i}$ and $\ket{k}$ respectively  in modes $\hat{a}$ and $\hat{b}$ and, after processing through the unitary $U$, measuring the Fock states $\ket{n}$ and $\ket{m}$ respectively  in modes $\hat{a}$ and $\hat{b}$.     \vspace{-0.4cm}  }
\end{figure}

The generating function (GF) of a sequence $\{ c_n \}_{n \in \mathbb{N}_0}$ is defined as
\begin{equation} \label{eq:GF}
g(z) \coloneqq \zt{n}{c_n}{z} = \sum_{n=0}^{\infty} c_n \, z^n , \qquad z\in\mathbb{C}.
\end{equation}
It is a powerful tool as $g(z)$ encapsulates the entire sequence via $c_n= n!^{-1}(\partial^n \! g/ \partial z^n)_{z=0}$.  Here, we exploit the properties of GFs in quantum optics when applied to the squared modulus of the matrix elements of Gaussian unitaries in Fock basis. Unlike the matrix elements in a coherent (Gaussian) basis, these happen to be quite difficult to handle because Fock states are non Gaussian, so it is helpful to characterize them via their GFs. Consider the 4-dimensional sequence of transition probabilities $|\bra{n,m}U\ket{i,k}|^2$ for some unitary $U$, where $\ket{i}$, $\ket{k}$, $\ket{n}$, and $\ket{m}$ denote Fock states ($i,k,n,m \in \mathbb{N}_0$). Its 4-variate GF can be written as (see \cite{SI})
\begin{equation}
f(\mathbf{v}) = \frac{\mathrm{Tr} \left[ (\tau_z \otimes \tau_w)U (\tau_x \otimes \tau_y) U^{\dagger} \right]}{(1-x)(1-y)(1-z)(1-w)},
\label{eq-f(v)}
\end{equation}
where we chose $\mathbf{v} \coloneqq (x,y,z,w)$ such that $(x,y) \in [0,1)^2$ and $(w,z) \in [0,1]^2$, with the conventions shown in Fig.~\ref{conventions}. Thus, $f(\mathbf{v})$ is proportional to the overlap between two Gaussian states, one of which being the product of two thermal states of the form  $\tau_x \coloneqq (1-x) \sum_{n=0}^{\infty} x^n \proj{n}$, while the other is the product of two thermal states processed through the unitary $U$. This makes $f(\mathbf{v})$ very easy to compute when $U$ is Gaussian, regardless of the complexity of $|\bra{n,m}U\ket{i,k}|^2$ itself, by exploiting the Gaussian  formalism in phase space. Recalling that the overlap between two zero-mean Gaussian states $\rho_1$ and $\rho_2$ with covariance matrices $V_1$ and $V_2$ is given by $\mathrm{Tr} [\rho_1 \rho_2] = 1/\sqrt{\det[(V_1+V_2)/2]}$~\cite{overlap}, the GF of $|\bra{n,m}\bs\ket{i,k}|^2$ can be expressed using standard tools of quantum optics as~\cite{SI}
\begin{equation}
f_{\eta}^{\mathrm{BS}}(\mathbf{v}) = \frac{1}{1 - \eta (xz+yw) - \bar{\eta} (xw+yz) + xyzw},
\label{f_eta^BS}
\end{equation}
where $\bar{\eta} \coloneqq 1-\eta$, while the GF of $|\bra{n,m}\tms\ket{i,k}|^2$ can be written as~\cite{SI}
\begin{equation}
f_{\lambda}^{\mathrm{TMS}}(\mathbf{v}) = \frac{\bar{\lambda}}{1 - \lambda (xy+zw) - \bar{\lambda} (xz+yw) + xyzw},
\label{f_eta^TMS}
\end{equation}
where $\bar{\lambda} \coloneqq 1-\lambda$. As a consistency check, we note that
\begin{eqnarray}
 && f_{\eta}^{\mathrm{BS}}(\mathbf{0}) = |\bra{0,0}\bs\ket{0,0}|^2  = 1 ,    \nonumber \\
 && f_{\lambda}^{\mathrm{TMS}}(\mathbf{0}) = |\bra{0,0}\tms\ket{0,0}|^2  = \bar{\lambda},
\end{eqnarray}
while normalization $\sum_{n,m=0}^\infty |\bra{n,m} U \ket{i,k}|^2 = 1, \forall i,k$, translates into
\begin{equation}
f^{\mathrm{BS/TMS}}(x,y,1,1) = (1-x)^{-1}(1-y)^{-1}.
\end{equation}
Interestingly, energy conservation in $\bs$ manifests itself through
$f_{\eta}^{\mathrm{BS}}(x,y,z,w)= f_{\eta}^{\mathrm{BS}}(tx,ty,z/t,w/t)$, $\forall t$,
while the conservation of the photon number difference in $\tms$ is reflected by $f_{\lambda}^{\mathrm{TMS}}(x,y,z,w)= f_{\lambda}^{\mathrm{TMS}}(tx,y/t,z/t,tw)$, $\forall t$, see~\cite{SI}.
\par

\begin{figure*}[t]
	\includegraphics[width=.8\paperwidth]{recBSdc.pdf}
	\caption{\label{recurrence-figure}Classical components of the recurrence formula~\eqref{eq:recBS} for the transition probability $B^{(i,k)}_{n}$ in a BS.}
\end{figure*}

\subsection{Multiphoton transition probabilities}
We use Eq.~\eqref{f_eta^BS} to derive a surprisingly simple recurrence equation for the multiphoton transition probabilities in a BS, denoted as
$B^{(i,k)}_{n} \coloneqq |\bra{n,i+k-n}\bs\ket{i,k}|^2$,
with $i,k,n \in \mathbb{N}_0$. Incidentally, note that a direct calculation yields~\cite{SI}
\begin{equation}
B_n^{(i,k)} = \eta^k \bar{\eta}^i \sum_{m,j=\max(0,n-k)}^{\min(i,n)} (-1)^{m+j} \, \gamma^{(i,k)}_{n,m,j} \left(\frac{\eta}{\bar{\eta}}\right)^{m+j-n},
\label{eq:complicated}
\end{equation}
where
\begin{equation}
    \gamma^{(i,k)}_{n,m,j} \coloneqq \binom{i}{m} \binom{k}{n-m} \binom{n}{j} \binom{i+k-n}{i-j},
\end{equation}
which is  quite cumbersome to manipulate. Nevertheless, the following theorem provides an alternative.
\begin{theo} \label{theo:mainBS}
If $i\!=\!k\!=\!n\!=\!0$, then $B_{n}^{(i,k)} = 1$, else,
\begin{align}
B_{n}^{(i,k)}  = & ~ \eta \, B_{n-1}^{(i-1,k)} + \eta \, B_{n}^{(i,k-1)}  \nonumber \\
& + \bar{\eta} \, B_{n}^{(i-1,k)} + \bar{\eta}  \, B_{n-1}^{(i,k-1)} - B_{n-1}^{(i-1,k-1)}.
\label{eq:recBS}
\end{align}
\end{theo}
The definition of $B^{(i,k)}_{n}$ is extended here to all integers $i,k,n$, setting it to zero when either of them is negative.

\textit{Proof.}
We set $\mathbf{u} \coloneqq (x,y,z)$, $\mathbf{j} \coloneqq (i,k,n)$ and denote by $g_{\eta}^{\mathrm{BS}}(\mathbf{u})$ the 3-variate GF of $B^{(i,k)}_{n}$ with the conventions of Fig.~\ref{conventions}.   
Since $g_{\eta}^{\mathrm{BS}}(\mathbf{u}) = f_{\eta}^{\mathrm{BS}}(x,y,z,1)$, Eq.~\eqref{f_eta^BS} implies
\begin{equation} \label{eq:relGFBS}
    \left[ 1 - \eta (xz+y) - \bar{\eta} (x+yz) + xyz \right] g_{\eta}^{\mathrm{BS}}(\mathbf{u}) = 1.
\end{equation}
Using the shifting property of the GFs and the notation of Eq.~\eqref{eq:GF}, it can easily be shown that multiplying the GF by $\mathbf{u}_l$ for $l=1,2,3$ corresponds to decreasing the index $\mathbf{j}_l$ of $B^{(i,k)}_{n}$ by one unit, so that for instance
\begin{equation}
    \zt{\mathbf{j}}{B_{n-1}^{(i-1,k)}}{\mathbf{u}} = xz \, \zt{\mathbf{j}}{B_{n}^{(i,k)}}{\mathbf{u}} = xz \, g_{\eta}^{\mathrm{BS}}(\mathbf{u}).
\end{equation}
In addition, we know that the 3-variate GF of the product $\delta_{i,0} \, \delta_{k,0} \, \delta_{n,0}$ of three Kronecker deltas is $1$.
Using this, we see that Eq. \eqref{eq:relGFBS} is equivalent to the relation
\begin{align}
& B_{n}^{(i,k)} - \eta \Big[ B_{n-1}^{(i-1,k)} + B_{n}^{(i,k-1)} \Big] \nonumber \\
& - \bar{\eta} \Big[ B_{n}^{(i-1,k)} + B_{n-1}^{(i,k-1)} \Big] + B_{n-1}^{(i-1,k-1)} = \delta_{i,0} \, \delta_{k,0} \, \delta_{n,0}, \nonumber
\end{align}
which proves the theorem. $\square$
\par

This recurrence can be nicely interpreted in the context of the HOM effect. As illustrated in Fig. \ref{recurrence-figure}, the first four terms of the right-hand side of Eq. \eqref{eq:recBS} corroborate the classical intuition one may have about $B_{n}^{(i,k)}$: one should add the probabilities corresponding to the different scenarios in which the $n$th photon has not reached the BS yet, multiplied by the right probability ($\eta$ or $\bar{\eta}$)  depending on which path it takes. For example, $B_{n-1}^{(i-1,k)}$ must be multiplied by $\eta$ since the extra photon must be injected on the input mode $a$ and exit on the output mode $a$. Crucially, as a consequence of bosonic statistics, a fifth term appears in Eq. \eqref{eq:recBS} with a minus sign that  accounts for quantum interference and may be viewed as an interference suppression term.
In the special case where $i=k=1$ and $\eta=1/2$, we recover the standard HOM effect,
\begin{align}
B_{1}^{(1,1)}  & =  \frac{1}{2} \, B_{0}^{(0,1)} \! + \frac{1}{2} \, B_{1}^{(1,0)} \! + \frac{1}{2} \, B_{1}^{(0,1)} \! + \frac{1}{2} \, B_{0}^{(1,0)} \! - B_{0}^{(0,0)}  \nonumber \\
& = 4 \times \frac{1}{2} \times \frac{1}{2} - 1 = 0 .
\label{eq:HOMrecurrence}
\end{align}
Let us stress that this is a very unconventional proof of the HOM effect as Eq.~\eqref{eq:HOMrecurrence} does not involve a linear combination of amplitudes but of probabilities. The first two terms account for both photons being transmitted while the third and fourth terms correspond to both of them being reflected. The fifth (negative) term has no classical counterpart. 
Note that if $k=0$ and $i\ge 0$, the interference term disappears in Eq.~\eqref{eq:recBS} and  one gets the recurrence $B^{(i,0)}_{n}= \eta \, B^{(i-1,0)}_{n-1} + \bar{\eta}  \, B^{(i-1,0)}_{n}$, which had been derived in the context of majorization theory applied to bosonic transformations~\cite{Christos}.
\par

\begin{figure*}[t]
	\includegraphics[width=.8\paperwidth]{recTMSdc.pdf}
	\caption{\label{recurrence-figure-2}Classical components of the recurrence formula~\eqref{eq:recTMS} for  the transition probabilities $A^{(i,k)}_{n} $ in a TMS.}
\end{figure*}

\subsection{Distinguishable photons}
It is instructive to give Eq.~\eqref{eq:recBS} further interpretation by juxtaposing it with its classical counterpart for distinguishable photons, which may for instance happen if the incident photons occupy different temporal modes. 
The classical probability of detecting $n$ photons on output mode $a$ when $i$ and $k$ photons impinge on input modes $a$ and $b$ is given by the convolution $p_{n|i,k}=\sum_{n'=0}^n p^A_{n'|i} \, p^B_{n-n'|k}$,   where $p^A_{n|i}$ (or $p^B_{n|k}$) is the probability of getting $n$ photons if $i$ (or $k$) distinguishable photons enter mode $a$ (or $b$), which itself follows a binomial distribution of parameter $\eta$ (or $\bar{\eta}$), see~\cite{SI} for details. Hence, the $3$-variate GF of $p_{n|i,k}$ is given by  $g_{\eta}^{\text{cl}}(\mathbf{u}) = g_{\eta}^{A}(x,z) \, g_{\eta}^{B}(y,z)$, where $g_{\eta}^{A}(x,z)$ and $g_{\eta}^{B}(y,z)$ are the $2$-variate GFs of $p^A_{n|i}$ and $p^B_{n|k}$. For instance, it is easy to show that $g_{\eta}^{A}(x,z) = 1/(1 - \eta xz - \bar{\eta} x)$, so that $g_{\eta}^{\text{cl}}(\mathbf{u})$ satisfies the relation
\begin{equation}
    \left( 1 - \eta xz - \bar{\eta} x \right) g_{\eta}^{\text{cl}}(\mathbf{u}) = 1(x) \, g_{\eta}^{B}(y,z)  .    
\label{eq:intermediate}
\end{equation}
where $1(x)\equiv 1$  is a constant function of $x$.
Using again the shifting property of GFs, Eq.~\eqref{eq:intermediate} implies the classical recurrence relation
\begin{equation} \label{eq:reccl1}
    p_{n|i,k} = \eta \, p_{n-1|i-1,k} + \bar{\eta} \, p_{n|i-1,k}, \qquad i>0,
\end{equation}
where we have used the fact that $1(x) \, g_{\eta}^{B}(y,z)$ is the GF of $\delta_{i,0} \, p^B_{n|k}$ and can be ignored for $i>0$.
Interchanging $p^A_{n|i}$ and $p^B_{n|k}$, a similar reasoning yields
\begin{equation} \label{eq:reccl2}
    p_{n|i,k} = \eta \, p_{n|i,k-1} + \bar{\eta} \, p_{n-1|i,k-1}, \qquad k>0.
\end{equation}
We notice here that Eq.~\eqref{eq:reccl1} coincides with the first and third terms in Eq.~\eqref{eq:recBS}, while Eq.~\eqref{eq:reccl2} coincides with the second and fourth terms. If either $i=0$ or $k=0$ (i.e., no photon in one of the two input modes), then Eq. \eqref{eq:recBS} reduces to the classical recurrence (for instance, Eq. \eqref{eq:reccl1} covers the case $k=0$). As advertised, the fifth (negative) term in Eq.~\eqref{eq:recBS} thus captures quantum interference (it appears as soon as $i,k>0$) since it is absent from the classical formulas \eqref{eq:reccl1} and \eqref{eq:reccl2}. Note also that removing this negative quantum term in Eq.~\eqref{eq:recBS} would then lead to twice the classical probability.

\par

\begin{figure}[b]
	\includegraphics[trim = 0cm 0cm 0cm 0cm, clip, width=.8\columnwidth]{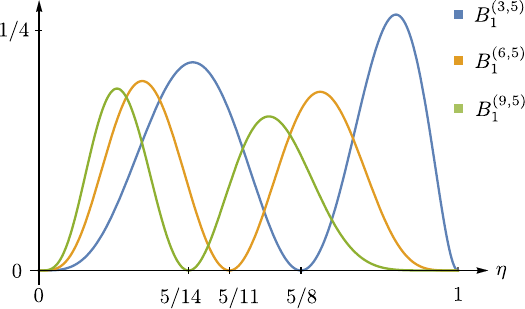}
	\caption{\label{rationalBS} Probability $B^{(i,k)}_1$ of observing a single photon on mode $a$ at the output of a BS for $k=5$ and for three different values $i = 3,6,9$ as a function of the transmittance $\eta$. We observe a HOM-like suppression effect for the corresponding rational values of $\eta = 5/8, 5/11, 5/14$.}
\end{figure}

\begin{figure}[b]
	\includegraphics[trim = 0cm 0cm 0cm 0cm, clip, width=.8\columnwidth]{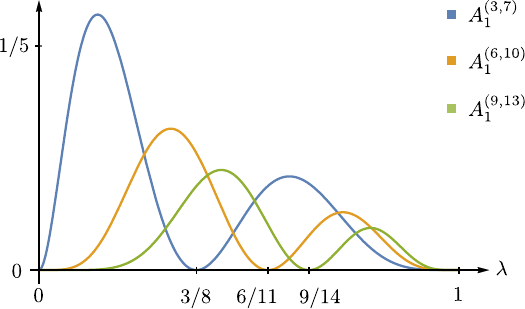}
	\caption{\label{rationalTMS} Probability $A^{(i,i+k-1)}_1$ of observing a single photon on mode $a$ at the output of a TMS for $k=5$ and for three different values $i = 3,6,9$ as a function of the parameter $\lambda$. Note that $k$ denotes here the number of output photons on mode~$b$. We observe a suppression effect akin to the HOM effect for the corresponding rational values of $\lambda = 3/8, 6/11, 9/14$, or equivalently for values of the gain $g = 8/5, 11/5, 14/5$.}
\end{figure}

\subsection{Active Gaussian transformations}
An even more appealing application of our framework is to explore multiphoton interferences in an active transformation, such as a TMS. As proven in \cite{HOM_PTR_2}, a TMS may be viewed as a BS undergoing ``partial time reversal'', namely $\bra{n,m} U^{\mathrm{TMS}}_{\lambda} \ket{i,k} = \sqrt{1-\lambda} \, \bra{n,k} U^{\mathrm{BS}}_{1-\lambda} \ket{i,m}$.
Indeed, indices $k$ and $m$ are interchanged, which may be interpreted as reverting the arrow of time of mode $b$~\cite{cewqo}. Similarly, interchanging variables $y$ and $w$, 
we see that the GFs are connected by $f_{\lambda}^{\mathrm{TMS}}(x,y,z,w) = (1-\lambda) \, f_{1-\lambda}^{\mathrm{BS}}(x,w,z,y)$, which is consistent with Eqs.~\eqref{f_eta^BS} and \eqref{f_eta^TMS}.
This allows us to write a recurrence for the transition probabilities $A^{(i,k)}_{n} \coloneqq |\bra{n,k-i+n}\tms\ket{i,k}|^2$ in a TMS (the definition of $A^{(i,k)}_{n}$ is extended to all integers $i,k,n$, setting it to zero when either of them is negative).
\begin{theo} \label{theo:mainTMS}
If $i=k=n=0$, then $A^{(i,k)}_{n} = \bar{\lambda}$, else,
\begin{align}
A_{n}^{(i,k)}  = & ~ \lambda \, A_{n}^{(i-1,k-1)} + \lambda \, A_{n-1}^{(i,k)}  \nonumber \\
& + \bar{\lambda} \, A_{n-1}^{(i-1,k)}  + \bar{\lambda} \, A_{n}^{(i,k-1)} - A_{n-1}^{(i-1,k-1)}.
\label{eq:recTMS}
\end{align}
\end{theo}
\textit{Proof.}
The relation can be easily proven  by making use of Theorem \ref{theo:mainBS}, exploiting the fact that $A^{(i,k)}_{n} = \bar{\lambda} \, B^{(i,k-i+n)}_{n}$ with $\eta=\bar{\lambda}$ (or $\eta=1/g$), see~\cite{SI}. $\square$

Equation~\eqref{eq:recTMS} is quite intriguing at first sight, as it is unclear how interferences take place in an active medium.
However, as illustrated in Fig. \ref{recurrence-figure-2}, we may build an interpretation of  Eq.~\eqref{eq:recTMS} by considering all possible classical scenarios.
The first term corresponds to the stimulated annihilation of an extra input photon pair, while the second term corresponds to the stimulated emission of an extra output photon pair (both occurring with probability $\propto \lambda$). The third and fourth terms correspond to an extra photon crossing the nonlinear medium without stimulating pair emission nor absorption (both with probability $\propto \bar{\lambda}$). Most importantly, the fifth (negative) term is again responsible for an unsuspected quantum interference effect, which has no classical counterpart. In the special case where $i=k=1$ and $\lambda=1/2$, we predict a complete extinction of the output state $\ket{1}\ket{1}$, which confirms a newly discovered two-photon interference effect in an amplifier of gain 2 \cite{HOM_PTR_2} originating from \emph{timelike} indistinguishability between the input and output photon pairs (exactly like the HOM effect can be viewed as a consequence of \emph{spacelike} indistinguishability between two photons entering a BS of transmittance 1/2). 
Here again, we find a surprising explanation of this effect based on the cancellation of probabilities (not amplitudes), namely
\begin{align}
A_{1}^{(1,1)}  & =  \frac{1}{2} \, A_{1}^{(0,0)} \! + \frac{1}{2} \, A_{0}^{(1,1)} \! + \frac{1}{2} \, A_{0}^{(0,1)} \! + \frac{1}{2} \, A_{1}^{(1,0)} \! - A_{0}^{(0,0)}  \nonumber \\
& = 4 \times \frac{1}{2} \times \frac{1}{4} - \frac{1}{2} = 0 .
\end{align}
The first two terms account for events consisting of the stimulated annihilation of the input photon pair accompanied with the stimulated emission of a distinct output pair, while the third and fourth terms correspond to events where both photons cross the TMS. The fifth term is intrinsically quantum.
Note that  for $k=0$, Eq. \eqref{eq:recTMS} reduces to the recurrence $A_{n}^{(i,0)}  =  \lambda \, A_{n-1}^{(i,0)} + \bar{\lambda} \, A_{n-1}^{(i-1,0)}$ implying a majorization relation in a bosonic amplifier channel that was proven in~\cite{Raul}.
\par

\subsection{Rational transmittance and gain}
Coming back to passive BS transformations, it is easy to predict the existence of a HOM-like suppression effect for any rational value of the transmittance $\eta < 1$ provided some specific numbers of impinging photons are considered, namely 
\begin{equation}
B^{(i,k)}_{1}=0 \qquad \mathrm{if~} \eta=k/(i+k),
\label{eq:extendedHOM-BS}
\end{equation}
as illustrated in Fig. \ref{rationalBS}. This can be understood as the result of amplitude cancellation between two scenarios, taking as a reference the situation where $i-1$ photons on mode $a$ are reflected and $k-1$ photons on mode $b$ are transmitted. The single photon observed on the output mode $a$ may come from input mode $a$ or $b$. Either the $i$th photon on mode $a$ is transmitted (there are $i$ possible choices) and all $k$ photons on mode $b$ are transmitted, which yields an amplitude $\propto i \, \eta$, or the $k$th photon on mode $b$ is reflected (there are $k$ possible choices) and all $i$ photons on mode $a$ are reflected, which yields an amplitude $\propto k \, \bar{\eta}$. Hence, we have $B_{1}^{(i,k)}  \propto  (i\, \eta - k\, \bar{\eta})^2$, which is consistent with Eq. \eqref{eq:extendedHOM-BS}. However, we provide a distinct interpretation in terms of probability cancellation as implied by Eq.~\eqref{eq:recBS}, see~\cite{SI}. 
For a quantum optical amplifier, we observe a similar effect for any rational value of the gain $g > 1$, namely
\begin{equation}
A^{(i,i+k-1)}_{1}=0 \qquad \mathrm{if~} \lambda=i/(i+k) ,
\label{eq:extendedHOM-TMS}
\end{equation}
corresponding to $g=1+i/k$ (see Fig. \ref{rationalTMS}). This heretofore unknown interference effect can again be viewed as a consequence of probability cancellation in Eq.~\eqref{eq:recTMS}, see~\cite{SI}.

\section{Conclusion and outlook}
Gaussian bosonic unitaries are readily described as affine transformations in phase space. Yet, addressing their action on Fock states typically leads to cumbersome calculations, which makes multiphotonic interferences in common Gaussian optical components hard to grasp. As a consequence, it is often an intractable task to prove fundamental entropy inequalities  for Gaussian bosonic channels, while these are of major importance in optical quantum communication (see, e.g., the entropy photon-number inequality  \cite{Guha2007,KonigSmith,DePalma2014}). Here, we have shown that the generating function of the matrix elements of a BS or TMS in Fock space can be expressed in a closed form, which, as a central consequence, yields  simple recurrence equations for the multiphoton transition probabilities.  In spite of the many interfering paths, Theorems~\ref{theo:mainBS} and \ref{theo:mainTMS} then provide a simple, intuitively appealing picture of multiphoton interference in passive and active bosonic circuits. It is amazing that such a simple account of quantum interferences in terms of probabilities (instead of amplitudes) in so well-studied optical components had yet remained unnoticed.

We have then predicted several multiphoton generalizations of the HOM effect in a BS of rational transmittance and have exploited the correspondence between a BS and TMS under  partial time reversal \cite{HOM_PTR_2} in order to reveal the existence of similar interferometric suppression effects in a quantum optical amplifier of rational gain. Interestingly, these predicted effects seem to escape the general framework for quantum suppression laws that has been derived in refs. \cite{Dittel-PRL2018,Dittel-PRA2018}.

Let us stress that the generating function of transition probabilities can also be useful in studying other properties of Gaussian unitaries, for example their asymptotic behavior. 
Using Tauberian theorems, which state that if $g(z) \sim 1/(1-z)$ for $z \rightarrow 1$, then $\sum_{l=0}^n c_l \sim n$ for $n \rightarrow \infty$,
it is indeed possible to approximate $B^{(i,i)}_{n}$ when $i \to \infty$ ~\cite{SI}. For $\eta=1/2$, this exactly coincides with the asymptotic analysis of a BS with a large photon number in both input ports~\cite{Saverio}. Note also that the generating function has recently been exploited in order to connect boson sampling with Fock-state inputs to boson sampling with thermal-state inputs \cite{Kim2020}, which is reminiscent of Eq.~\eqref{eq-f(v)}. Moreover, the technique developed here yields a powerful tool for characterizing certain non-Gaussian bosonic channels (those that are Gaussian-dilatable), for example photon-added or photon-subtracted channels~\cite{Carlos2012,Krishna2017,PhysRevA.98.013809} as well as the linear coupling of a signal mode together with a passive environment~\cite{FockMajPassChannels}.

Overall, beyond the results for a BS and TMS highlighted in this paper, we expect that our framework can be amenable to address any Bogoliubov transformation acting on an arbitrary number of modes. The special case of a multimode linear interferometer has already been considered in \cite{Jabbour-PhD,Jabbour-next}. Although it does not seem to have implications for the complexity of simulating bosonic interferences, it provides a neat description of multimode multiphoton interference involving negative probabilities. Furthermore, we may anticipate other applications of this framework going beyond photonic systems. The same approach should indeed prove valuable for nonphotonic bosonic systems as well, since the transformations described by Eqs.~\eqref{eq:BogoliubovBS} and \eqref{eq:BogoliubovTMS} are not restricted to optical components but have quite a broad range of applications. In short, we have at hand a distinct approach to quantum multiparticle interferences in (passive and active) Bogoliubov transformations acting on any bosonic quantum systems.



M.G.J. acknowledges support from the Wiener-Anspach Foundation and from the Carlsberg Foundation. This work was supported by the F.R.S.-FNRS under project no. PDR T.0224.18 and by the EC under project ShoQC within the ERA-NET Cofund in Quantum Technologies (QuantERA) program.


\bibliography{BiblioQO_GaussianUnitaries_arXiv}


\onecolumngrid

\appendix

\section*{Supplemental Material}

The present material provides supplementary information to the main text. It includes calculations which are standard in both quantum optics and the theory of generating functions, which we chose not to cover in the main document. Section \ref{sec:ProbaBS} provides the calculation of the explicit expression of the multiphoton transition probabilities in a beam splitter, which is used as a benchmark to exhibit the interest of our developed framework. Section \ref{sec:GenFunc} covers the calculation of the generating functions of these probabilities as well as of the corresponding probabilities for a two-mode squeezer (whose explicit form we chose not to include as its complexity makes it of little interest). The recurrence equations on the multiphoton transition probabilities that can be deduced from these generating functions (see main text) are then discussed in the case of a beam splitter with a rational transmittance or two-mode squeezer with a rational gain.
The calculation of the classical transition probabilities for distinguishable photons impinging on a beam splitter, which serves as a reference in the analysis of our main results, is summarized in Section \ref{sec:Probacl}. Finally, Section \ref{sec:Asymp} explores the asymptotic behavior of the transition probabilities, illustrating the power of generating functions.

\subsection{Multiphoton transition probabilities in a beam splitter} \label{sec:ProbaBS}

Consider a beam splitter (BS) of transmittance $\eta \in [0,1]$ characterized by the unitary $\bs$ of the form
\begin{equation}
    \bs = \exp \left[ \theta \left( \ad \b - \a \bd \right) \right], \quad \eta = \cos^2 \theta.
\end{equation}
An expression for the transition amplitudes $b^{(i,k)}_n \coloneqq \bra{n,m}\bs\ket{i,k}$ (for $i,k,n \in \mathbb{N}_0$ and $m=i+k-n$) can be computed by first deriving an expression for the following state:
\begin{equation}
    \ket{\psi^{(i,k)}} = \bs \ket{i,k} = \frac{1}{\sqrt{k!}} \bs (\bd)^k \bsd \bs \ket{i,0} = \frac{1}{\sqrt{k!}} \left( \bs \bd \bsd \right)^k \ket{\psi^{(i,0)}}.
\end{equation}
Exploiting the action of the BS in phase space, we get
\begin{equation}
\ket{\psi^{(i,k)}} = \frac{1}{\sqrt{k!}} \left(\sqrt{1-\eta} \ad + \sqrt{\eta} \bd \right)^k \ket{\psi^{(i,0)}} = \frac{1}{\sqrt{k!}} \sum_{m=0}^k {k \choose m} \left(\sqrt{1-\eta} \ad \right)^m \left( \sqrt{\eta} \bd \right)^{k-m} \ket{\psi^{(i,0)}}.
\label{Psiik1}
\end{equation}
Similarly, we have
\begin{equation}
\ket{\psi^{(i,0)}} = \frac{1}{\sqrt{i!}} \left( \sqrt{\eta} \ad - \sqrt{1-\eta} \bd \right)^i \ket{0,0} = \sum_{n=0}^i (-1)^{i-n} \sqrt{{i \choose n} \eta^n (1-\eta)^{i-n}} \ket{n,i-n}.
\label{Psii0}
\end{equation}
Combining Eqs. (\ref{Psiik1}) and (\ref{Psii0}), we obtain
\begin{equation}
\ket{\psi^{(i,k)}} = \sum_{n,m=0}^{i,k} (-1)^{i-n} \sqrt{\Gamma^{(i,k)}_{n,m} \eta^{n+k-m} (1-\eta)^{i-n+m}} \ket{n+m,i-n+k-m},
\label{Psiik2}
\end{equation}
where we defined
\begin{equation}
\Gamma^{(i,k)}_{n,m} = {i \choose n} {k \choose m} {n+m \choose n} {i-n+k-m \choose i-n}.
\end{equation}
Form this, we obtain
\begin{equation}
    b_n^{(i,k)} = \sum_{m=\max(0,n-k)}^{\min(i,n)} (-1)^{i-m} \sqrt{\Gamma^{(i,k)}_{m,n-m} \eta^{2m+k-n} (1-\eta)^{i-2m+n}},
\end{equation}
where
\begin{equation}
    \Gamma^{(i,k)}_{n,m} \coloneqq \binom{i}{n} \binom{k}{m} \binom{n+m}{n} \binom{i-n+k-m}{i-n}.
\end{equation}
Thus, the transition probabilities $B^{(i,k)}_{n} \coloneqq \left| b_n^{(i,k)}\right|^2$ in the BS $\bs$ are given by
\begin{equation}
    B_n^{(i,k)}  = \sum_{m,j=\max(0,n-k)}^{\min(i,n)} (-1)^{m+j} \sqrt{\Gamma^{(i,k)}_{m,n-m} \Gamma^{(i,k)}_{j,n-j}} \eta^{k-n+m+j} (1-\eta)^{i+n-m-j}.
    \label{Bikn1}
\end{equation}
Now, it can easily be shown that
\begin{equation}
    \Gamma^{(i,k)}_{m,n-m} \Gamma^{(i,k)}_{j,n-j} = \gamma^{(i,k)}_{n,m,j} \gamma^{(i,k)}_{n,j,m},
\end{equation}
where we defined
\begin{equation}
    \gamma^{(i,k)}_{n,m,j} \coloneqq \binom{i}{m} \binom{k}{n-m} \binom{n}{j} \binom{i+k-n}{i-j}.
\end{equation}
Since $\gamma^{(i,k)}_{n,m,j} = \gamma^{(i,k)}_{n,j,m}$, we end up with $\Gamma^{(i,k)}_{m,n-m} \Gamma^{(i,k)}_{j,n-j} = (\gamma^{(i,k)}_{n,m,j})^2$, so that
\begin{equation}
B_n^{(i,k)} = \sum_{m,j=\max(0,n-k)}^{\min(i,n)} (-1)^{m+j} \gamma^{(i,k)}_{n,m,j} \eta^{k-n+m+j} (1-\eta)^{i+n-m-j}.
\label{Bikn2}
\end{equation}
This expression can of course be evaluated, but it is rather cumbersome and hence not very useful for the analytical investigation of multiphoton interferometry in a beam splitter. For instance, it is written as a double summation over terms with alternating signs, so that the positivity of the transition probability is not obvious from the expression. Similar derivations have for instance been given in \cite{Kim2002},\cite{Krishna2011} and \cite{Krishna2017}.

\subsection{Generating functions of the multiphoton transition probabilities} \label{sec:GenFunc}

\subsubsection{Case of a beam splitter}

The generating function (GF) of the transition probability $|\bra{n,m} \bs \ket{i,k}|^2$ in a BS is given by a function $f_{\eta}^{\mathrm{BS}} : [0,1)^2 \times [0,1]^2 \rightarrow [0,\infty)$ defined as
\begin{equation} \label{eq:defGenFuncProbaBS}
f_{\eta}^{\mathrm{BS}}(x,y,z,w) \coloneqq \sum_{i,k,n,m} |\bra{n,m} \bs \ket{i,k}|^2 x^i y^k z^n w^m,
\end{equation}
where, when we omit limits in summations, it means that the summation is carried out over all natural numbers in $\mathbb{N}_0$ (including $0$). The trick is to realize that it can rewritten as
\eqaln{
f_{\eta}^{\mathrm{BS}}(x,y,z,w) & = \mathrm{Tr} \left[ \bs \left( \sum_i x^i \proj{i} \otimes \sum_k y^k \proj{k} \right) \bsd \left( \sum_n z^n \proj{n} \otimes \sum_m w^m \proj{m} \right) \right] \\
& = \frac{1}{(1-x) (1-y) (1-z) (1-w)} \mathrm{Tr} \left[ \bs \left( \tau_x \otimes \tau_y \right) \bsd \left( \tau_z \otimes \tau_w \right) \right]
}
where $\tau_t$ is a Gaussian thermal state of parameter $t$~\cite{Weedbrook2011}, i.e,
\begin{equation}
\tau_t \coloneqq (1-t) \sum_m t^m \proj{m}.
\end{equation}
Now, the object $\rho_1 \coloneqq \bs \left( \tau_x \otimes \tau_y \right) \bsd$ actually represents the effect of a beam-splitter unitary on the tensor product of two Gaussian thermal states, making it a two-mode Gaussian state. The object $\rho_2 \coloneqq \tau_z \otimes \tau_w$ is obviously a two-mode Gaussian state as well. This means that $f_{\eta}^{\mathrm{BS}}(x,y,z,w)$ is proportional to the overlap $\text{Tr} \left[ \rho_1 \rho_2 \right]$ between the two Gaussian states $\rho_1$ and $\rho_2$,
\begin{equation}
f_{\eta}^{\mathrm{BS}}(x,y,z,w) = \frac{1}{(1-x) (1-y) (1-z) (1-w)} \text{Tr} \left[ \rho_1 \rho_2 \right].
\end{equation}
The above quantity can therefore be computed easily using standard tools of Gaussian quantum optics, \textit{i.e.}, the symplectic formalism applied to the phase-space representation of bosonic quantum systems~\cite{Weedbrook2011}. Since the first moments of each of the two Gaussian states $\rho_1$ and $\rho_2$ is zero, their overlap can be computed using the formula~\cite{overlap}
\begin{equation}
\text{Tr} \left[ \rho_1 \rho_2 \right] = \left(\text{det}\left[ \frac{V_1 + V_2}{2} \right]\right)^{-\frac{1}{2}} = \frac{4}{\sqrt{\text{det}\left[ V_1 + V_2 \right]}},
\end{equation}
where $V_1$ and $V_2$ are the respective covariance matrices of $\rho_1$ and $\rho_2$.
Some easy matrix algebra involving covariance matrices and the symplectic matrix of the BS in phase space finally yields
\begin{equation}
f_{\eta}^{\mathrm{BS}}(x,y,z,w) = \frac{1}{1 - \eta xz - (1-\eta) xw - \eta yw - (1-\eta) yz + xyzw}.
\label{OGFProbBS}
\end{equation}

The conservation of energy in the BS can be easily verified using the GF given by the above equation. Define the function $\tilde{f}_{\eta}^{\mathrm{BS}} : [0,1)^2 \times [0,1]^3 \rightarrow [0,\infty)$ as
\begin{equation}
\tilde{f}_{\eta}^{\mathrm{BS}}(x,y,z,w,t) \coloneqq \sum_{i,k,n,m} |\bra{n,m} \bs \ket{i,k}|^2 x^i y^k z^n w^m t^{i+k-n-m} = f_{\eta}^{\mathrm{BS}}(xt,yt,\frac{z}{t},\frac{w}{t}).
\label{tildegBS}
\end{equation}
From Eq. \eqref{OGFProbBS}, we have
\begin{equation}
\tilde{f}_{\eta}^{\mathrm{BS}}(x,y,z,w,t) = f_{\eta}^{\mathrm{BS}}(x,y,z,w), \quad \forall t.
\end{equation}
This actually means that $\tilde{f}_{\eta}^{\mathrm{BS}}$ as defined in Eq. \eqref{tildegBS} does not depend on variable $t$, so that the only non-zero elements in the sums of the right-hand side of Eq. \eqref{tildegBS} verify $i+k-n-m=0$. Consequently,
\begin{equation}
\bra{n,m} \bs \ket{i,k} = 0 \quad \mathrm{if} \quad i+k \neq n+m.
\end{equation}

\subsubsection{Symmetric inputs to the beam splitter}
We now consider the case in which the same Fock states impinge on both inputs of the BS and compute the GF of the corresponding transition probabilities, which will be useful when investigating the asymptotic behavior of $B^{(i,i)}_n$ (see Section \ref{sec:Asymp}). The sequence $B^{(i,k)}_n$ depends on 3 indices only, index $m$ in $|\bra{n,m} \bs \ket{i,k}|^2$ being redundant as a consequence of energy conservation in a BS. The GF of $B^{(i,k)}_n$ is then simply given by
\eqaln{
\zt{i,k,n}{B^{(i,k)}_n}{x,y,z} & \coloneqq \sum_{i,k,n} B^{(i,k)}_n x^i y^k z^n \\
& = \sum_{i,k,n} \left( \sum_m |\bra{n,m} \bs \ket{i,k}|^2 \right) x^i y^k z^n \\
& = f_{\eta}^{\mathrm{BS}}(x,y,z,1).
}
In order to derive the GF of the diagonal elements $B^{(i,i)}_n$, we force the relation $k=i$ in the GF of $B^{(i,k)}_n$ by only considering the elements which satisfy it. Using the notation $c_n = [z^n] g(z)$ to mean that we select the coefficient of the $z^n$ term in $g(z) \coloneqq \sum_{n=0}^{\infty} c_n z^n$, we write
\begin{equation}
    \zt{i,n}{B^{(i,i)}_n}{x,z} = [y^0] \sum_{i,k,n} B^{(i,k)}_n x^i y^{k-i} z^n = [y^0] f_{\eta}^{\mathrm{BS}}\left(\frac{x}{y},y,z,1\right).
\end{equation}
By Cauchy's integral formula for any function $g(z)$, one has
\begin{equation}
g(a) = \frac{1}{2 \pi i} \oint \mathrm{d} z \frac{g(z)}{z-a}.
\end{equation}
Applying this to our case, we get that, for some circle $\gamma_x$ around $y=0$,
\begin{equation}
\zt{i,n}{B^{(i,i)}_n}{x,z} = \frac{1}{2 \pi i} \int_{\gamma_x} \mathrm{d} y \frac{f_{\eta}^{\mathrm{BS}}\left(x/y,y,z,1\right)}{y}.
\end{equation}
Now, using the Residue Theorem, the above equation amounts to
\begin{equation}
\zt{i,n}{B^{(i,i)}_n}{x,z} = \sum_l \mathrm{Res} \left[ \frac{f_{\eta}^{\mathrm{BS}}\left(x/y,y,z,1\right)}{y} ; y = s_l(x,z,\eta) \right],
\end{equation}
where the $s_l$ represent the singularities of $f_{\eta}^{\mathrm{BS}}\left(x/y,y,z,1\right)/y$ satisfying $\lim_{x \rightarrow 0} s_l(x,z,\eta) = 0$.
Some standard calculations yield
\begin{equation}
s_{1,2}(x,z,\eta) = \frac{1+xz \pm \sqrt{(1+xz)^2-4(\eta +(1-\eta)z)(x(1-\eta)+\eta x z))}}{2(\eta+z(1-\eta))},
\end{equation}
with the subscript $1$($2$) corresponding to $+$($-$) in the $\pm$ in the above equation. If we take their limits for $x$ approaching zero, we obtain
\begin{equation}
\lim_{x \rightarrow 0} s_1(x,z,\eta) = 0 \quad \mathrm{and} \quad \lim_{x \rightarrow 0} s_2(x,z,\eta) = \frac{1}{\eta + z(1-\eta)}.
\end{equation}
The residue of the function we are interested in reduces to
\begin{equation}
\mathrm{Res} \left[ \frac{f_{\eta}^{\mathrm{BS}}\left(x/y,y,z,1\right)}{y} ; y = s_1(x,z,\eta) \right] = \frac{1}{\sqrt{(1+xz)^2-4(\eta +(1-\eta)z)(x(1-\eta)+\eta x z))}},
\end{equation}
so that
\begin{equation}
\zt{i,n}{B^{(i,i)}_n}{x,z} = \frac{1}{\sqrt{(1+xz)^2-4(\eta +(1-\eta)z)(x(1-\eta)+\eta x z))}}.
\end{equation}
If we particularize this to a balanced BS ($\eta=1/2$), we obtain the simple expression
\begin{equation} \label{eq:GFdiagBalanced}
\zt{i,n}{\left. B^{(i,i)}_n \right|_{\eta=1/2}}{x,z} = \frac{1}{\sqrt{(1-x)(1-z^2 x)}},
\end{equation}
which is the GF in $i,n$ of the diagonal sequence $B^{(i,i)}_n$ for $\eta = 1/2$.
This will be useful for analyzing the asymptotic behavior of the transition probabilities in Section \ref{sec:Asymp}.

\subsubsection{Case of a two-mode squeezer}

The GF of the transition probability $|\bra{n,m} \tms \ket{i,k}|^2$ in a two-mode squeezer (TMS) is given by a function $f_{\lambda}^{\mathrm{TMS}} : [0,1)^2 \times [0,1]^2 \rightarrow [0,\infty)$ defined as
\begin{equation} \label{eq:defGenFuncProbaTMS}
f_{\lambda}^{\mathrm{TMS}}(x,y,z,w) \coloneqq \sum_{i,k,n,m} |\bra{n,m} \tms \ket{i,k}|^2 x^i y^k z^n w^m,
\end{equation}
with $\lambda=\tanh^2 (r)$. One could compute it from scratch similarly as in the previous section. A much more elegant option is to use a fundamental relation that links the BS and the TMS, which was proven in~\cite{HOM_PTR_2}. There, it was shown that the TMS may be viewed as a BS undergoing ``partial time reversal''~\cite{cewqo},  $|\bra{n,m} U^{\mathrm{TMS}}_{\lambda} \ket{i,k}|^2 = (1-\lambda) \, |\bra{n,k} U^{\mathrm{BS}}_{1-\lambda} \ket{i,m}|^2$. As explained in the main text, it implies that the GFs are connected by the relation
\begin{equation}
    f_{\lambda}^{\mathrm{TMS}}(x,y,z,w) = (1-\lambda) \, f_{1-\lambda}^{\mathrm{BS}}(x,w,z,y),
\end{equation}
which then yields
\begin{equation}
f_{\lambda}^{\mathrm{TMS}}(x,y,z,w) = \frac{1-\lambda}{1 - \lambda (xy+zw) - (1-\lambda) (xz+yw) + xyzw}.
\label{f_eta^TMS_SI}
\end{equation}

\subsubsection{Rational value of the transmittance or gain}

We now discuss the interferometric suppression effect that exists for a BS of rational transmittance $\eta<1$ or for a TMS of rational gain $g>1$. It can be checked that
\begin{equation}
B_{1}^{(i,k)}  = 0 \qquad \mathrm{if~}\eta=\frac{k}{i+k}, \qquad \mathrm{where~} i,k\in \mathbb{N}_0, i+k > 0,
\label{eq:B1ik}
\end{equation}
which gives rise a to full interferometric suppression for any rational $\eta<1$, extending the HOM effect for $i=k=1$ and $\eta=1/2$. Indeed, using the mode transformation characterizing the BS (see main text), a closed expression for $B_{1}^{(i,k)} $ can easily be written, which entails the sum of the amplitudes where either a single photon from mode $a$ is  transmitted (all $k$ photons on mode $b$ being transmitted) or a single photon from mode $b$ is reflected (all $i$ photons on mode $a$ being reflected) weighted with the appropriate combinatorial factors, namely 
\begin{equation}
\label{eq:closed_B1ik}
B_{1}^{(i,k)}  =  \frac {i+k-1}{i\, k} \,  (i \, \eta - k \, \bar{\eta})^2\, B_{0}^{(i-1,k-1)}  \, ,
\end{equation}
where we have used the trivial expression 
\begin{equation}
\label{eq:closed_B0ik}
B_{0}^{(i,k)} = \binom{i+k}{i}\, \bar{\eta}^i \, \eta^k  \, .
\end{equation}
Note that given the symmetry between the two modes as well as the input-output symmetry, we have three associated interferometric suppressions (having in common a single photon in either one of the input or output mode):
\begin{equation}
    \begin{aligned}
        & B_{i+k-1}^{(i,k)} = 0 && \mathrm{if~}\eta=\frac{i}{i+k}, i+k > 0,\\
        & B_{n}^{(1,k+n-1)} = 0 \qquad && \mathrm{if~}\eta=\frac{k}{k+n}, k+n>0, \\
        & B_{n}^{(i+n-1,1)} = 0 && \mathrm{if~}\eta=\frac{n}{n+i}, i+n>0.
    \end{aligned}
\end{equation}
It is instructive to examine the interference effect \eqref{eq:B1ik} by using the recurrence equation for a BS derived in the main text, yielding
\begin{equation}
B_{1}^{(i,k)}  =  \eta \, B_{0}^{(i-1,k)} + \eta \, B_{1}^{(i,k-1)}  + \bar{\eta} \, B_{1}^{(i-1,k)} + \bar{\eta}  \, B_{0}^{(i,k-1)} - B_{0}^{(i-1,k-1)}. 
\label{eq:recur_for_B1ik}
\end{equation}
Note first that applying the recurrence equation to $B_{0}^{(i,k)}$ instead of $B_{1}^{(i,k)}$ gives
\begin{equation}
B_{0}^{(i,k)}  =   \eta \, B_{0}^{(i,k-1)}  + \bar{\eta} \, B_{0}^{(i-1,k)} \, ,
\label{eq:Pascal1}
\end{equation}
which, using Eq. \eqref{eq:closed_B0ik}, simply reduces to Pascal's formula for binomial coefficients
\begin{equation}
\binom{i+k}{i} = \binom{i+k-1}{i} + \binom{i+k-1}{i-1}  \, .
\label{eq:Pascal2}
\end{equation}
Thus, as expected, the recurrence  equation \eqref{eq:Pascal1} is classical and can be recovered using simple combinatorial analysis. By comparison, Eq.~\eqref{eq:recur_for_B1ik} gives a more interesting recurrence. Using the above closed expression for $B_{0}^{(i,k)}$, it can be rewritten as
\begin{equation}
B_{1}^{(i,k)}  = \eta \, B_{1}^{(i,k-1)}  + \bar{\eta} \, B_{1}^{(i-1,k)}-\kappa \, B_{0}^{(i-1,k-1)} , \qquad \mathrm{with~} \kappa= 1-(i+k-1) \left(\frac{\eta^2}{k}+\frac{\bar{\eta}^2}{i}\right).\label{eq:recur_for_B1ik_2nd}
\end{equation}
The closed expression \eqref{eq:closed_B1ik} is of course a solution of Eq. \eqref{eq:recur_for_B1ik_2nd}, but we see that the recurrence here is more interesting as it exhibits again  a negative probability term.
If we choose $\eta=k/(i+k)$ (with $i+k>0$), then $\kappa=1/(i+k)$, so the interferometric suppression effect $B_{1}^{(i,k)}  = 0$ translates Eq. \eqref{eq:recur_for_B1ik_2nd} into
\begin{equation}
k \, B_{1}^{(i,k-1)}  + i \, B_{1}^{(i-1,k)} - B_{0}^{(i-1,k-1)}  = 0 \, .
\end{equation}
The first and second terms can be interpreted classically : starting from the case where the output mode $a$ already contains a single photon, the first term accounts for the $k$th photon in mode $b$ being transmitted (so it does not give an extra photon on output mode $a$) while the second term accounts for the $i$th photon in mode $a$ being reflected  (so it does not give an extra photon on output mode $a$). Remarkably, the third (negative) term has no classical meaning and results in the full cancellation of probability $B_{1}^{(i,k)} $.

The same analysis can be applied to a quantum optical amplifier. As shown in the main text, we have
\begin{equation}
A^{(1,k)}_{n}=0 \qquad \mathrm{if~} \lambda=\frac{n}{n+k} \mathrm{~or~}  g=1+\frac{n}{k}, \qquad \mathrm{where~} n,k\in \mathbb{N}_0, k>0,
\label{eq:recur_for_An1k}
\end{equation}
which gives rise to full a interferometric suppression for any rational $\lambda<1$ (or rational gain $g=1/(1-\lambda)>1$). 
When $k=n=1$ and $\lambda=1/2$, we confirm the existence of an interferometric suppression effect in a parametric amplifier of gain $g=2$ \cite{HOM_PTR_2}. 
Similarly as for a BS, we also have three associated interferometric suppressions in a TMS, namely 
\begin{equation}
    \begin{aligned}
        & A^{(i,1)}_{i+n-1}=0 && \mathrm{if~} \lambda=\frac{n}{i+n} \mathrm{~or~}  g=1+\frac{n}{i}, i>0, \\
        & A^{(i,i+k-1)}_{1}=0 && \mathrm{if~} \lambda=\frac{i}{i+k} \mathrm{~or~}  g=1+\frac{k}{i}, i>0, \\
        & A^{(k+n-1,k)}_{n}=0 \qquad && \mathrm{if~} \lambda=\frac{k}{k+n} \mathrm{~or~}  g=1+\frac{k}{n}, n>0.
    \end{aligned}    
\end{equation}
Let us examine the interference Eq. \eqref{eq:recur_for_An1k} by using the recurrence equation for a TMS, 
\begin{equation}
A_{n}^{(1,k)}  =  \lambda \, A_{n}^{(0,k-1)} + \lambda \, A_{n-1}^{(1,k)}  
 + \bar{\lambda} \, A_{n-1}^{(0,k)}  + \bar{\lambda} \, A_{n}^{(1,k-1)} - A_{n-1}^{(0,k-1)}.
\end{equation}
Using the closed expression
\begin{equation}
A_{n}^{(0,k)}  = (1-\lambda) \, B_{0}^{(n,k)} = \binom{n+k}{k} \lambda^n \bar{\lambda}^{k+1},
\end{equation}
we may reexpress it as
\begin{equation}
A_{n}^{(1,k)}  =   \lambda \, A_{n-1}^{(1,k)}  + \bar{\lambda} \, A_{n}^{(1,k-1)} 
 - \kappa' A_{n-1}^{(0,k-1)} , \qquad \mathrm{with~} \kappa'= 1-(n+k-1) \left(\frac{\lambda^2}{n}+\frac{\bar{\lambda}^2}{k}\right).
\end{equation}
Similarly as for a BS, if we choose $\lambda=n/(n+k)$ (or $g=1+n/k$, with $k>0$),  the interferometric suppression $A^{(1,k)}_{n}=0$ implies 
\begin{equation}
 n \, A_{n-1}^{(1,k)}  + k \, A_{n}^{(1,k-1)}  -  A_{n-1}^{(0,k-1)} =0
\end{equation}
Here, taking as a reference the situation where the input mode $a$ contains a single photon, the first term accounts for the stimulated emission of a photon pair at the output (with probability $\propto \lambda$) while the second term accounts for the single photon in mode $b$ being transmitted (with probability  $\propto \bar{\lambda}$).
Again, the third (negative) term has no classical interpretation and is responsible for the cancellation $A_{n}^{(1,k)}=0$.

\subsection{Multiphoton transition probabilities in a beam splitter with distinguishable photons} \label{sec:Probacl}



Consider a situation in which the photons impinging on the two input modes $a$ and $b$ of the BS are distinguishable. The incident photons may for instance have different polarizations. We now count the photons exiting the BS in mode $a'$, without making a distinction between different photons. The fact that they are distinguishable will however affect the distributions of photons in the output modes. The probability $\mathbb{P}(n \text{ in } a'| i \text{ in } a)$ that we detect $n$ photons in output mode $a'$ when we sent $i$ photons in mode $a$ is given by a simple binomial of parameter $\eta$, \textit{i.e.},
\begin{equation} \label{eq:seq1}
    p_a(n|i) \coloneqq \mathbb{P}(n \text{ in } a'| i \text{ in } a) = \binom{i}{n} \eta^n (1-\eta)^{i-n}.
\end{equation}
Similarly, the probability $\mathbb{P}(n \text{ in } a'| k \text{ in } b)$ that we detect $n$ photons in mode $a'$ when we sent $k$ photons in mode $b$ is given by
\begin{equation} \label{eq:seq2}
    p_b(n|k) \coloneqq \mathbb{P}(n \text{ in } a'| k \text{ in } b) = \binom{k}{n} (1-\eta)^n \eta^{k-n}.
\end{equation}
Using this, the probability $p(n|i,k) \coloneqq \mathbb{P}(n \text{ in } a'| i \text{ in } a \text{ and } k \text{ in } b)$ that we detect $n$ photons in mode $a'$ when we sent $i$ photons in mode $a$ and $k$ photons in mode $b$ can be calculated using a convolution
\begin{equation}
p(n|i,k)
= \sum_{n'=0}^n p_a(n'|i) p_b(n-n'|k).
\label{classBS1}
\end{equation}
The $3$-variate GF of the sequence $p(n|i,k)$ is given by a function $g_{\eta}^{\text{cl}} : [0,1)^2 \times [0,1] \rightarrow [0,\infty)$ defined as
\begin{equation}
g_{\eta}^{\text{cl}}(x,y,z) \coloneqq \zt{i,k,n}{p(n|i,k)}{x,y,z} = \sum_{i,k,n} p(n|i,k) x^i y^k z^n.
\end{equation}
Since the sequence $p(n|i,k)$ is given by a convolution over index $n$ of the sequences $p_a(n|i)$ and $p_b(n|k)$, the GF $g_{\eta}^{\text{cl}}(x,y,z)$ is simply given by the product of their two respective GFs $g_{\eta}^{a} : [0,1) \times [0,1]$ and $g_{\eta}^{b} : [0,1) \times [0,1]$, which can simply be computed as
\begin{equation} \label{eq:GFAB}
g_{\eta}^{a}(x,z) = \frac{1}{1 - \eta xz - (1-\eta) x}, \qquad g_{\eta}^{b}(y,z) = \frac{1}{1 - \eta y - (1-\eta) yz}.
\end{equation}
This means that $g_{\eta}^{\text{cl}}$ satisfies the relation
\begin{equation} \label{eq:relGFcl}
    \left[ 1 - \eta xz - (1-\eta) x \right] g_{\eta}^{\text{cl}}(x,y,z) = g_{\eta}^{b}(y,z) = \sum_{i=0}^{\infty} g_{\eta}^{b}(y,z) \delta_{i0} x^i,
\end{equation}
where $\delta_{\cdot \cdot}$ denotes a Kronecker delta. Using the shifting property of the GFs, the counterpart of Eq. \eqref{eq:relGFcl} for sequences is
\begin{equation} \label{eq:reccl1_SI}
    p(n|i,k) - \eta \; p(n-1|i-1,k) - \bar{\eta} \; p(n|i-1,k) = \delta_{i0} \, p_a(n|k),
\end{equation}
where $\bar{\eta} = 1-\eta$  and $p(n|i,k) = 0$ if either of the indices $n,i,k$ is negative. A similar reasoning yields
\begin{equation} \label{eq:reccl2_SI}
    p(n|i,k) - \bar{\eta}\; p(n-1|i,k-1) - \eta \; p(n|i,k-1) = \delta_{k0} \, p_b(n|i).
\end{equation}
Obviously, by summing the two relations one can always write the weaker relation
\eqal{ \label{eq:reccl3}
    p(n|i,k) & = \nu \left[ \eta \, p(n-1|i-1,k) + \bar{\eta} \, p(n|i-1,k) + \delta_{i0} \, p_b(n|k) \right] \\
    & \quad + (1-\nu) \left[ \bar{\eta} \, p(n-1|i,k-1) + \eta \, p(n|i,k-1) + \delta_{k0} \, p_a(n|i) \right]
}
for any $\nu \in [0,1]$, which amounts to
\begin{equation}
    p(n|i,k) =\begin{cases}
    1, & \text{ if } i=0 \text{ and } k=0, \vspace{.2cm} \\
    \nu \, p_b(n|k) + (1-\nu) \left[ \bar{\eta} \, p(n-1|i,k-1) + \eta \, p(n|i,k-1) \right], & \text{ if } i=0 \text{ and } k \neq 0, \vspace{.2cm} \\
    \nu \left[ \eta \, p(n-1|i-1,k) + \bar{\eta} \, p(n|i-1,k) \right] + (1-\nu) \, p_a(n|i), & \text{ if } i \neq 0 \text{ and } k = 0, \vspace{.2cm} \\
    \nu \left[ \eta \, p(n-1|i-1,k) + \bar{\eta} \, p(n|i-1,k) \right] \\
     \quad + (1-\nu) \left[ \bar{\eta} \, p(n-1|i,k-1) + \eta \, p(n|i,k-1) \right], & \text{ else}.
    \end{cases}
\end{equation}
The last relation can be written more simply as
\begin{equation}
    p(n|i,k) =\begin{cases}
    1, & \text{ if } i=0 \text{ and } k=0, \vspace{.2cm} \\
    p_b(n|k), & \text{ if } i=0 \text{ and } k \neq 0, \vspace{.2cm} \\
    p_a(n|i), & \text{ if } i \neq 0 \text{ and } k = 0, \vspace{.2cm} \\
    \nu \left[ \eta \, p(n-1|i-1,k) + \bar{\eta} \, p(n|i-1,k) \right] \\
     \quad + (1-\nu) \left[ \bar{\eta} \, p(n-1|i,k-1) + \eta \, p(n|i,k-1) \right], & \text{ else}.
    \end{cases}
\end{equation}

\subsection{Asymptotics of the transition probabilities} \label{sec:Asymp}

The asymptotic behavior of a sequence $\left\lbrace c_n \right\rbrace$ for a growing index can be studied by analyzing the asymptotic behavior of the corresponding GF $g(z)$ around its singularities. This is encompassed in the Tauberian theorems~\cite{Flajolet}, the most famous of which being due to Hardy, Littlewood~\cite{Tauberian1} and Karamata~\cite{Tauberian2}.~\\

\noindent\textbf{The HLK Tauberian theorem.} { \it Let $g(z)$ be a power series with radius of convergence equal to $1$, satisfying
\begin{equation}
g(z) \sim \frac{1}{(1-z)^{\alpha}} \Lambda\left(\frac{1}{1-z}\right), \quad z \rightarrow 1,
\end{equation}
for some $\alpha \geq 0$ with $\Lambda$ a slowly varying function. Assume that the coefficients $c_n = [z^n] g(z)$ are all non-negative. Then
\begin{equation}
\sum_{k=0}^n c_k \sim \frac{n^{\alpha}}{\Gamma(\alpha+1)} \Lambda(n), \quad n \rightarrow \infty.
\end{equation}
}
A function $\Lambda$ is said to be slowly varying at infinity if and only if, for any $\beta > 0$, one has
\begin{equation}
\frac{\Lambda(\beta x)}{\Lambda(x)} \rightarrow 1 \quad \mathrm{as} \quad x \rightarrow + \infty.
\end{equation}

Our aim is now to use Tauberian theorems in order to study the asymptotic behavior of $B^{(i,i)}_n$ for $\eta = 1/2$. The HLK Tauberian theorem can be generalized, and in case of multiple singularities, each one can be analyzed separately, and the different contributions can be combined in the end~\cite{Flajolet}. In our case, this must be done in two steps, since our sequence has two indices $i$ and $n$. We begin by analyzing the behavior of
\begin{equation}
[z^n] \zt{i,n}{\left. B^{(i,i)}_n \right|_{\eta=1/2}}{x,z} = \zt{i}{\left. B^{(i,i)}_n \right|_{\eta=1/2}}{x},
\end{equation}
the GF in $i$, by studying the behavior of
\begin{equation}
\zt{i,n}{\left. B^{(i,i)}_n \right|_{\eta=1/2}}{x,z},
\end{equation}
the GF in $i$ and $n$. We then investigate the resulting
\begin{equation}
\zt{i}{\left. B^{(i,i)}_n \right|_{\eta=1/2}}{x}
\end{equation}
in order to conclude about
\begin{equation}
\left. B^{(i,i)}_n \right|_{\eta=1/2}.
\end{equation}

\noindent\textbf{Behavior of \texorpdfstring{$\zt{i}{\left. B^{(i,i)}_n \right|_{\eta=1/2}}{x}$}{}.}
The function given in Eq. \eqref{eq:GFdiagBalanced} has two singularities $z_1(x) \coloneqq 1/\sqrt{x}$ and $z_2(x) \coloneqq -1/\sqrt{x}$.
First,
\begin{equation}
\zt{i,n}{\left. B^{(i,i)}_n \right|_{\eta=1/2}}{x,z} \sim \frac{1}{\sqrt{2(1-x)(1-\sqrt{x} z)}}, \quad \text{when } z \rightarrow z_1(x).
\end{equation}
Define the sequence $\beta^{(1)}_{i,n}$ such that
\begin{equation}
\sum_{i,n} \beta^{(1)}_{i,n} x^i z^n = \frac{1}{\sqrt{2(1-x)(1-\sqrt{x} z)}}.
\label{eqAsymp1}
\end{equation}
In other words,
\begin{equation}
\zt{i,n}{\left. B^{(i,i)}_n \right|_{\eta=1/2}}{x,z} \sim \sum_{i,n} \beta^{(1)}_{i,n} x^i z^n, \quad \text{when } z \rightarrow z_1(x).
\end{equation}
Equation (\ref{eqAsymp1}) is the same as ($x$ is positive)
\begin{equation}
\sum_{i,n} \beta^{(1)}_{i,n} x^i \left(\frac{z}{\sqrt{x}}\right)^n = \frac{1}{\sqrt{2(1-x)(1-z)}},
\end{equation}
or,
\begin{equation}
\sum_{i,n} \beta^{(1)}_{i,n} x^{i-\frac{n}{2}} z^n = \frac{1}{\sqrt{2(1-x)(1-z)}}.
\end{equation}
Now, for $n$ increasing, according to the Tauberian theorems,
\begin{equation}
[z^n] \frac{1}{\sqrt{2(1-x)(1-z)}} \sim \frac{1}{\sqrt{2(1-x) \pi n}},
\end{equation}
so that
\begin{equation}
[z^n] \sum_{i,n} \beta^{(1)}_{i,n} x^{i-\frac{n}{2}} z^n \sim \frac{1}{\sqrt{2(1-x) \pi n}},
\end{equation}
\begin{equation}
[z^n] \sum_{i,n} \beta^{(1)}_{i,n} x^i z^n \sim \frac{x^{\frac{n}{2}}}{\sqrt{2(1-x) \pi n}}.
\end{equation}
Using Definition (\ref{eqAsymp1}), we end up with
\begin{equation}
[z^n] \frac{1}{\sqrt{2(1-x)(1-\sqrt{x} z)}} \sim \frac{x^{\frac{n}{2}}}{\sqrt{2(1-x) \pi n}}.
\end{equation}
Secondly,
\begin{equation}
\zt{i,n}{\left. B^{(i,i)}_n \right|_{\eta=1/2}}{x,z} \sim \frac{1}{\sqrt{2(1-x)(1+\sqrt{x} z)}}, \quad \text{when } z \rightarrow z_2(x).
\end{equation}
We can do the same analysis, and obtain
\begin{equation}
[z^n] \frac{1}{\sqrt{2(1-x)(1+\sqrt{x} z)}} \sim \frac{(-1)^n x^{\frac{n}{2}}}{\sqrt{2(1-x) \pi n}}.
\end{equation}
As we explained earlier, in the case of two singularities (having the same absolute value), the two asymptotic contributions can be added up~\cite{Flajolet}, so that
\begin{equation}
[z^n] \zt{i,n}{\left. B^{(i,i)}_n \right|_{\eta=1/2}}{x,z} \sim \frac{x^{\frac{n}{2}}}{\sqrt{2(1-x) \pi n}} + \frac{(-1)^n x^{\frac{n}{2}}}{\sqrt{2(1-x) \pi n}},
\end{equation}
or,
\begin{equation} \label{eq:AsympGFdiagBalanced}
\zt{i}{\left. B^{(i,i)}_n \right|_{\eta=1/2}}{x} \sim \frac{1+(-1)^n}{\sqrt{2\pi n}} \frac{x^{\frac{n}{2}}}{\sqrt{1-x}}.
\end{equation}
The zero contribution for odd $n$ is consistent with the fact that the total input photon number $2i$ is even.~\\

\noindent\textbf{Behavior of \texorpdfstring{$\left. B^{(i,i)}_n \right|_{\eta=1/2}$}{}.}
The function on the right-hand side of Eq. \eqref{eq:AsympGFdiagBalanced} has only one singularity, $x_0 \coloneqq 1$. Since the dominant factor is $1/\sqrt{1-x}$ (compared to $x^{\frac{n}{2}}$) when $x \rightarrow x_0$, we can focus on it. We have~\cite{Flajolet}
\begin{equation}
[x^i] \frac{1}{\sqrt{1-x}} \sim \frac{1}{\sqrt{\pi i}},
\end{equation}
meaning that
\begin{equation}
[x^i] \left( x^{- \frac{n}{2}} \zt{i}{\left. B^{(i,i)}_n \right|_{\eta=1/2}}{x} \right) \sim \frac{1+(-1)^n}{\sqrt{2\pi n}} \frac{1}{\sqrt{\pi i}}.
\label{eqAsymp2}
\end{equation}
Now,
\eqaln{
x^{- \frac{n}{2}} \zt{i}{\left. B^{(i,i)}_n \right|_{\eta=1/2}}{x} & = x^{- \frac{n}{2}} \sum_{i=0}^{\infty} \left. B^{(i,i)}_n \right|_{\eta=1/2} x^i \\
& = \sum_{i=0}^{\infty} \left. B^{(i,i)}_n \right|_{\eta=1/2} x^{i- \frac{n}{2}} \\
& = \sum_{j=-n/2}^{\infty} \left. B^{(j+\frac{n}{2},j+\frac{n}{2})}_n \right|_{\eta=1/2} x^j,
}
and $B^{(i,i)}_n = 0$ if $n>2i$, so that
\begin{equation}
x^{- \frac{n}{2}} \zt{i}{\left. B^{(i,i)}_n \right|_{\eta=1/2}}{x} = \sum_{j=0}^{\infty} \left. B^{(j+\frac{n}{2},j+\frac{n}{2})}_n \right|_{\eta=1/2} x^j,
\end{equation}
and
\begin{equation}
[x^i] \left( x^{- \frac{n}{2}} \zt{i}{\left. B^{(i,i)}_n \right|_{\eta=1/2}}{x} \right) = \left. B^{(i+\frac{n}{2},i+\frac{n}{2})}_n \right|_{\eta=1/2}.
\end{equation}
As a consequence of Equation (\ref{eqAsymp2}),
\begin{equation}
\left. B^{(i+\frac{n}{2},i+\frac{n}{2})}_n \right|_{\eta=1/2} \sim \frac{1+(-1)^n}{\sqrt{2\pi n}} \frac{1}{\sqrt{\pi i}},
\end{equation}
or,
\begin{equation}
\left. B^{(i,i)}_n \right|_{\eta=1/2} \sim \frac{1+(-1)^n}{\sqrt{2\pi n}} \frac{1}{\sqrt{\pi \left(i-\frac{n}{2}\right)}}.
\end{equation}
After some simplification, we obtain
\begin{equation}
\left. B^{(i,i)}_n \right|_{\eta=1/2} \sim \frac{1+(-1)^n}{\pi \sqrt{n \left(2i-n\right)}}.
\end{equation}
which exactly coincides with the result of the analysis performed in~\cite{Saverio}. The output terms around $n \sim i$ are maximally suppressed, which is reminiscent of the HOM effect. Interestingly, we can again exploit partial time reversal and extend this analysis to a TMS with $\lambda=1/2$, giving
\begin{equation}
A^{(i,k)}_k \sim \frac{1+(-1)^i}{2\pi \sqrt{i(2k-i)}}, \quad k,i \rightarrow \infty .
\end{equation}

\end{document}